# Multiple equilibrium states of a curved-sided hexagram: Part I—Stability of states[1]

Lu Lu[a], Jize Dai[a], Sophie Leanza[a], Ruike Renee Zhao[a], John W. Hutchinson[b,*]

[a] Department of Mechanical Engineering, Stanford University, Stanford, CA 94305, USA
[b] School of Engineering and Applied Sciences, Harvard University, Cambridge, MA 02138, USA

*Corresponding author. Email: jhutchin@fas.harvard.edu

**Abstract**

The stability of the multiple equilibrium states of a hexagram ring with six curved sides is investigated. Each of the six segments is a rod having the same length and uniform natural curvature. These rods are bent uniformly in the plane of the hexagram into equal arcs of 120º or 240º and joined at a cusp where their ends meet to form a 1-loop planar ring. The 1-loop rings formed from 120º or 240º arcs are inversions of one another and they, in turn, can be folded into a 3-loop straight line configuration or a 3-loop ring with each loop in an "8" shape. Each of these four equilibrium states has a uniform bending moment. Two additional intriguing planar shapes, 6-circle hexagrams, with equilibrium states that are also uniform bending, are identified and analyzed for stability. Stability is lost when the natural curvature falls outside the upper and lower limits in the form of a bifurcation mode involving coupled out-of-plane deflection and torsion of the rod segments. Conditions for stability, or lack thereof, depend on the geometry of the rod cross-section as well as its natural curvature. Rods with circular and rectangular cross-sections will be analyzed using a specialized form of Kirchhoff rod theory, and properties will be detailed such that all four of the states of interest are mutually stable. Experimental demonstrations of the various states and their stability are presented. Part II presents numerical simulations of transitions between states using both rod theory and a three-dimensional finite element formulation, includes confirmation of the stability limits established in Part I, and presents additional experimental demonstrations and verifications.

---

# 1. Introduction

Nonlinear elastic solids and structures having multiple equilibrium states are of interest in themselves and because of their utility for a variety of functional applications (Lachenal et al., 2012, Mouthuy et al., 2012, Olson et al, 2013, Mhatre et al., 2021, Leanza et al, 2022, Wu et al., 2022). In this paper, a six-sided ring, which will be referred to as a curved-sided hexagram, is investigated which can assume an interesting array of simple planar equilibrium states, including a state in which the ring folds into a collapsed state in the form of a straight 3-loop configuration. Part I of the paper investigates the conditions for stability of some of the most interesting states using a specialized formulation of Kirchhoff rod theory along with a limited set of experimental verifications. Part II investigates aspects of the transition from one state to the other and presents more extensive experiments illustrating the relevant phenomena. Four of the curved-sided hexagram's equilibrium states of interest are shown in Fig. 1. These states are planar without out-of-plane deformation, as demonstrated in our experiments shown in Fig. A1 in Appendix A. Experimental details on how to fabricate these curved-sided hexagram rings are provided in Fig. A2 in Appendix B. Any one of these states might be regarded as the fundamental ring structure, but they are all simply different states of the same entity. We begin by discussing the star hexagram.

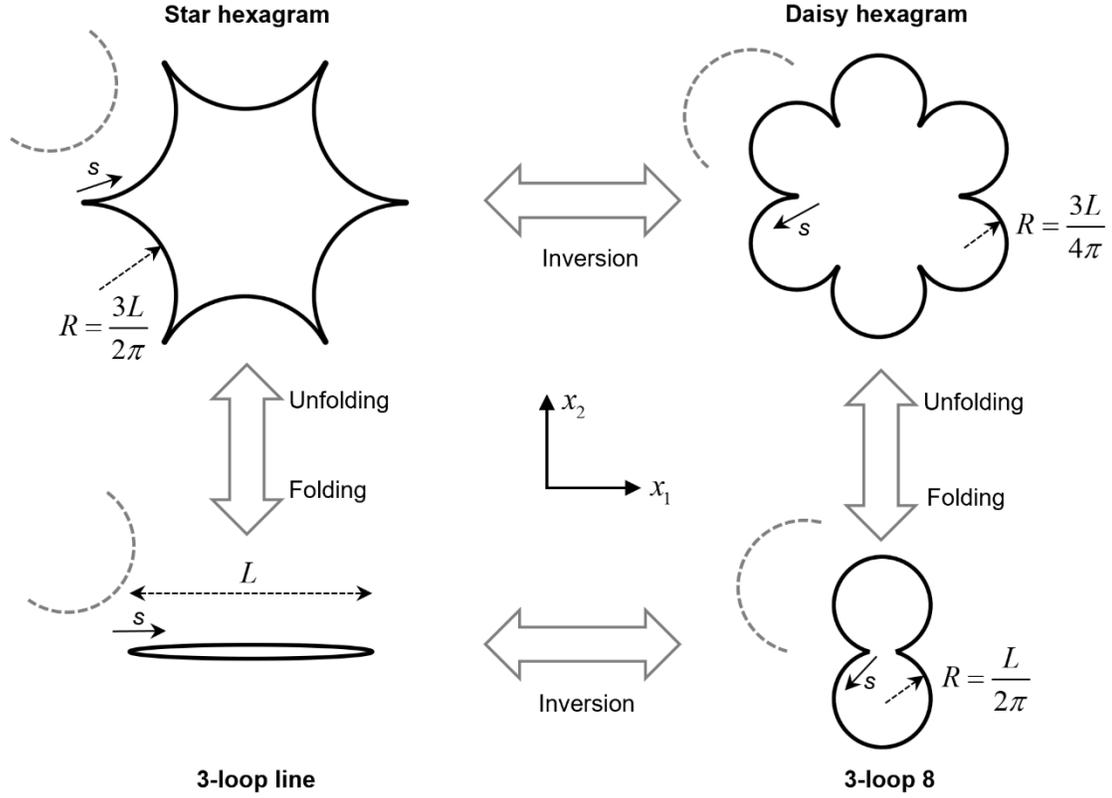

Fig. 1. Four equilibrium states of the curved-sided hexagram showing the origin of the distance measure and the direction traversing the ring corresponding to increasing $s$ used in assigning numbers to the segments. The entities are planar and lie in the $(x_1, x_2)$ plane. The three loops in the two states on the lower half of the figure are stacked on top of one another. The dashed curves depict a positive natural curvature $\kappa_n$ (about the axis $\mathbf{i}_3$) of the detached arcs as seen by an observer traversing the ring in the direction indicated. The length of each rod is $L$ and, for the three states with curved segments, the radius $R$ of the arc is given in terms of $L$. In the equilibrium state each segment carries a uniform bending moment about $\mathbf{i}_3$, $B_3(R^{-1} - \kappa_n)$, becoming $-B_3\kappa_n$ for the 3-loop line state. The flip of the dashed arcs, depicting a positive natural curvature, from the states on the left to those on the right occurs due to inversion.

To construct the star hexagram, one begins with six uniform rods of length $L$, each having a uniform natural curvature $\kappa_n$ in the plane in which the hexagram is formed (perpendicular to $\mathbf{i}_3$), with the positive sense indicated by the dashed arc in the upper left-hand portion of Fig. 1. Naturally straight rods have $\kappa_n = 0$. Each rod is bent about the $\mathbf{i}_3$-axis to obtained a uniform radius of curvature $R = 3L/2\pi$ in the plane of curved hexagram, corresponding to a 120° arc, and then the six arc segments are 'welded' at the six cusps where their ends meet resulting in the planar curved-sided ring shown in Fig. 1. The constitutive bending and torsional behavior of the rods in this paper will be taken as linear with $B_1$ as the out-of-plane stiffness and $B_3$ as the in-plane stiffness about rod axes in the planar state of Fig. 1. The torsional stiffness about the rod centerline is denoted by $B_2$. In the equilibrium state each rod segment carries a uniform bending moment of magnitude $B_3(R^{-1}-\kappa_n)$ about the $\mathbf{i}_3$-axis consistent with local and overall equilibrium of the ring with no externally applied forces or moments.

We will demonstrate by simulation and experiment that the star hexagram will transform to the daisy hexagram, also shown in Fig. 1, with a proper choice of external stimuli. The daisy hexagram is an inverted star hexagram and vice versa. Inverted here means turned 'inside-out' with the inner surface of one state becoming the outer surface of the other state and vice versa. The inversion process involves out-of-plane behavior with bending and twisting, but the inverted ring is planar. Alternatively, one could imagine constructing the daisy hexagram from scratch in a manner similar to that described for the star hexagram. For the daisy, the positive convention for the natural curvature flips from that for the star (because of inversion) as depicted in Fig. 1, and the radius of curvature to which each rod is bent (with sign consistent with increasing $s$) is now $R = 3L/4\pi$ such that the rod is a 240° arc.

We will also demonstrate the transformation of the star hexagram to the 3-loop line state and the transformation of the daisy hexagram to the 3-loop 8 state, and vice versa. The two 3-loop states consist of 3 stacked pairs of rod segments with each pair forming either two rods side by side in the line state or a planar configuration "8" with cusps meeting at the center. Each of the four equilibrium states in Fig. 1 has a uniform bending moment in each rod equal to $B_3(R^{-1}-\kappa_n)$ where the radius $R$ depends on the state ($R^{-1}=0$ for the line state). There will be cross-over of the rods from one planar level of the stack to another involving slight bending and torsion, but in the one-dimensional rod model this departure from pure bending is neglected. The strain energy in the ring in the equilibrium state is $3B_3L(R^{-1}-\kappa_n)^2$.

While the simplicity of the uniform bending moment in each of the four states in Fig. 1 makes equilibrium almost obvious, it is not obvious which, if any, of these states will be stable and therefore observable. Establishing the stability, or lack thereof, of the four states in Fig. 1 is the objective of Part I of this paper. Two other uniform bending states corresponding to 6-circle hexagrams will also be introduced and analyzed for stability. Some preliminaries follow.

*1.1. Preliminaries of the stability analysis*

The stability analysis used here is a specialized version of Kirchhoff rod theory developed in Leanza et al. (2023). Displacements from the equilibrium state and Euler angles are both employed. In the reference equilibrium state, each rod segment of the ring under investigation, lies in the $(x_1, x_2)$ plane and has a uniform radius, $R$, which will be infinite if the segment is straight. There is no twist of the rod segments in any of the initial equilibrium states considered in this paper. The Euler angles, $(\alpha, \beta, \gamma)$, defined in Fig. 2, measure angular differences from unit cylindrical base vectors in the reference state, $(\mathbf{i}_r, \mathbf{i}_\theta, \mathbf{i}_3)$, and the triad of unit base vectors attached to and rotating with the rod centerline, $(\mathbf{e}_1, \mathbf{e}_2, \mathbf{e}_3)$, with $\mathbf{e}_2$ tangent to the centerline in the direction increasing distance $s$. In the reference state, $(\mathbf{e}_1, \mathbf{e}_2, \mathbf{e}_3)$ coincide with $(\mathbf{i}_r, \mathbf{i}_\theta, \mathbf{i}_3)$ and the Euler angles are zero because there is no initial twist. The displacement components measured from the reference state are $\mathbf{u}(s) = u_r(s)\mathbf{i}_r + u_\theta(s)\mathbf{i}_\theta + u_3(s)\mathbf{i}_3$ in planar polar coordinates, where $s$ increases with distance along the rod centerline.

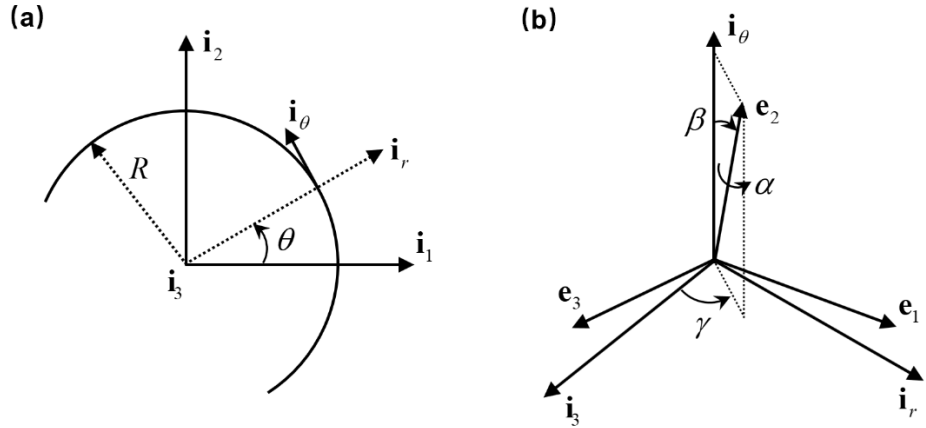

Fig. 2. Unit vectors and Euler angles. a) Cartesian base vectors and cylindrical base vectors in the reference state. b) Local cylindrical base vector triad and the Euler angles, $\alpha$, $\beta$ and $\gamma$, specifying the three embedded unit base vectors, $\mathbf{e}_i$.

The centerline of the rod is taken to be inextensional from which it follows that

$$\varepsilon_\theta = \sqrt{1-\varphi_r^2 - \varphi_3^2} - 1, \tag{1.1}$$

with the linearized stretching strain, $\varepsilon_\theta$, and rotations, $\varphi_r$ and $\varphi_3$, of the centerline defined as

$$\varepsilon_\theta = (du_\theta/ds + R^{-1}u_r), \ \varphi_r = du_3/ds, \ \varphi_3 = -(du_r/ds - R^{-1}u_\theta). \tag{1.2}$$

The displacement gradients are related to the Euler angles by (Leanza et al., 2023)

$$\cos\beta = \sqrt{1-\varphi_r^2 - \varphi_3^2}, \ \varphi_r = \sin\beta\cos\gamma, \ \varphi_3 = -\sin\beta\sin\gamma. \tag{1.3}$$

If $\sin\beta \neq 0$, it also follows that

$$\tan\gamma = -\varphi_3/\varphi_r. \tag{1.4}$$

The expressions above are exact, as are the following components of the vector, $\Omega = \kappa_1 \mathbf{e}_1 + \kappa_2 \mathbf{e}_2 + \kappa_3 \mathbf{e}_3$, measuring curvature of the centerline,

$$\begin{aligned}
\kappa_1 &= \cos\alpha\,(d\beta/ds - R^{-1}\sin\gamma) - \sin\alpha(-\sin\beta\,d\gamma/ds + R^{-1}\cos\beta\cos\gamma), \\
\kappa_2 &= d\alpha/ds + \cos\beta\,d\gamma/ds + R^{-1}\sin\beta\cos\gamma, \\
\kappa_3 &= \cos\alpha\,(-\sin\beta\,d\gamma/ds + R^{-1}\cos\beta\cos\gamma) + \sin\alpha(d\beta/ds - R^{-1}\sin\gamma).
\end{aligned} \tag{1.5}$$

The curvature components are expressed in terms of the unit vector triad embedded in the cross-section of the rod with $\kappa_2$ as the twist (rotation per distance) of the cross-section about its centerline. The elastic strain energy in the deformed state in a rod segment for any of the problems considered in this paper is

$$SE = \frac{1}{2}\int_0^L \left\{ B_1 \kappa_1^2 + B_2 \kappa_2^2 + B_3 (\kappa_3 - \kappa_n)^2 \right\} ds. \tag{1.6}$$

## 2. Stability analysis of the equilibrium states

For admissible displacements from any one of the four basic states under consideration, the strain energy in the curved hexagram in the deformed state is

$$SE = \frac{1}{2} \sum_{i=1}^{6} \int_{(i-1)L}^{iL} \left\{ B_1 \kappa_1^{(i)2} + B_2 \kappa_2^{(i)2} + B_3 \left( \kappa_3^{(i)} - \kappa_n \right)^2 \right\} ds, \qquad (2.1)$$

with the index $i$ labeling the rod segments traversing around the ring in the direction of increasing $s$ starting at $s = 0$ as indicated in Fig. 1. In each of the basic states, the ring lies in the $(x_1, x_2)$ plane and the curvature, $\kappa_1$, and twist, $\kappa_2$, are zero while $\kappa_3 = R^{-1}$. In the previous paper (Leanza, et al., 2023), the quadratic functional governing the change in strain energy of the ring, (2.1), from this basic state has been derived which is accurate to second order in the displacements and their gradients. This functional is the second variation of the energy of the ring and it governs stability of the basic state (Leanza, et al., 2023):

$$P_2(\alpha, \beta, \kappa_n) = \frac{1}{2} \sum_{i=1}^{6} \int_{(i-1)L}^{iL} \left\{ B_1 \left( \frac{d\beta^{(i)}}{ds} - \frac{\alpha^{(i)}}{R} \right)^2 + B_2 \left( \frac{d\alpha^{(i)}}{ds} + \frac{\beta^{(i)}}{R} \right)^2 - B_3 M \left( \left( \frac{\beta^{(i)}}{R} \right)^2 + \left( \frac{\alpha^{(i)}}{R} \right)^2 + 2 \frac{d\alpha^{(i)}}{ds} \frac{\beta^{(i)}}{R} \right) \right\} ds.$$

(2.2)

The dimensionless parameter measuring the bending moment in the basic state, $M = 1 - R\kappa_n$, was introduced by Audoly and Seffen (2015) in their extensive study of circular ring stability and behavior. If $P_2 > 0$ for all non-zero admissible displacements, excluding rigid body motion, the state is stable. If $P_2 < 0$ for any admissible displacement field that is not a rigid body motion, the state is unstable. To the order of accuracy required of the functional, $\gamma = 0$, implying that the in-plane displacements, $u_r$ and $u_\theta$, are not involved, and $du_3/ds = \beta$. We will replace $\beta$ in (2.2) by $du_3/ds$ to avoid having to enforce a constraint on $\beta$ associated with the requirement that $u_3$ must be continuous around the ring. Inextensionality imposes no additional constraint to this order. The functional remains valid for the straight sided 3-loop ring with $1/R \to 0$.

The dimensionless natural curvature $R\kappa_n$, or equivalently, $M$, determines stability. If $R\kappa_n = 1$ ($M = 0$), $P_2 \geq 0$ for all non-trivial admissible displacements, and only vanishes for rigid body motions. We seek the range of natural curvatures, $R\kappa_n^- < R\kappa_n < R\kappa_n^+$, where $R\kappa_n^- < 1$ and $R\kappa_n^+ > 1$, such that $P_2 > 0$ for all non-trivial admissible displacements, excluding rigid body motion. The limits of the range, $\kappa_n^-$ and $\kappa_n^+$, are such that a non-zero admissible displacement field which is not a rigid body motion exists delivering $P_2 = 0$. That admissible displacement field, or fields, is called the bifurcation mode, or modes. The problem of determining the limiting values of the stable range of $\kappa_n$ and the associated mode or modes is the eigenvalue problem addressed in this section associated with finding solutions that render $P_2$ stationary.

It will be convenient to work with dimensionless quantities, and for that purpose let

$$x = (s/L - (i-1)), \ (\ )' = d(\ )/dx, \ u = u_3/L, \ \bar{L} = L/R, \ b_1 = B_1/B_3, \ b_2 = B_2/B_3, \quad (2.3)$$

with $\alpha$ and $u$ now regarded as functions of $x$, with $0 \le x \le 1$ in each segment. The stability functional becomes

$$\frac{P_2(\alpha, u, R\kappa_n)}{B_3 L^{-1}} = \frac{1}{2} \sum_{i=1}^{6} \int_0^1 \left\{ b_1 \left( u^{(i)''} - \bar{L}\alpha^{(i)} \right)^2 + b_2 \left( \alpha^{(i)'} + \bar{L}u^{(i)'} \right)^2 - M\bar{L}\left( \bar{L}u^{(i)'2} + \bar{L}\alpha^{(i)2} + 2\alpha^{(i)'}u^{(i)'} \right) \right\} dx.$$

(2.4)

The three geometric conditions linking two rods joined at a common cusp are

$$u^{(i+1)}(0) = u^{(i)}(1), \ u^{(i+1)'}(0) = -u^{(i)'}(1), \ \alpha^{(i+1)}(0) = -\alpha^{(i)}(1), \quad \text{for } i = 1, 6, \quad (2.5)$$

with the understanding that for $i = 6$, $i+1$ becomes 1. Note that the direction along the ring with increasing $s$ changes by 180° at the cusp, and this accounts for the change in sign of $u'$ and $\alpha$ across the cusp.

The sixth order ODE system rendering $P_2$ stationary requires that in each segment

$$\begin{aligned} b_1(u'''' - \bar{L}\alpha'') - b_2\bar{L}(\alpha'' + \bar{L}u'') + M\bar{L}(\bar{L}u'' + \alpha'') &= 0, \\ b_1\bar{L}(u'' - \bar{L}\alpha) + b_2(\alpha'' + \bar{L}u'') + M\bar{L}(\bar{L}\alpha - u'') &= 0. \end{aligned} \quad (2.6)$$

If $(b_1 - M)(b_2 - M) > 0$ the general solution to these equations in segment $i$ is

$$\begin{aligned} u^{(i)} &= c_1^{(i)} + c_2^{(i)} \cos(\bar{L}x) + c_3^{(i)} \cos(\lambda \bar{L}x) + c_4^{(i)} x + c_5^{(i)} \sin(\bar{L}x) + c_6^{(i)} \sin(\lambda \bar{L}x), \\ \alpha &= -\bar{L}\left( c_2^{(i)} \cos(\bar{L}x) + c_3^{(i)} q \cos(\lambda \bar{L}x) + c_5^{(i)} \sin(\bar{L}x) + c_6^{(i)} q \sin(\lambda \bar{L}x) \right), \end{aligned} \quad (2.7)$$

with $\lambda = \sqrt{(b_1 - M)(b_2 - M)/b_1 b_2}$, while if $(b_1 - M)(b_2 - M) < 0$ it is

$$u^{(i)} = c_1^{(i)} + c_2^{(i)} \cos(\bar{L}x) + c_3^{(i)} \cosh(\lambda \bar{L}x) + c_4^{(i)} x + c_5^{(i)} \sin(\bar{L}x) + c_6^{(i)} \sinh(\lambda \bar{L}x),$$
$$\alpha = -\bar{L}\left(c_2^{(i)} \cos(\bar{L}x) + c_3^{(i)} q \cosh(\lambda \bar{L}x) + c_5^{(i)} \sin(\bar{L}x) + c_6^{(i)} q \sinh(\lambda \bar{L}x)\right),$$
(2.8)

with $\lambda = \sqrt{-(b_1 - M)(b_2 - M)/b_1 b_2}$. In both cases, $q = (b_2 - M)/b_2$. The natural, or dynamic, conditions at the joints complementing the geometric conditions in (2.5) are

$$u^{(i+1)\prime\prime}(0) = -u^{(i)\prime\prime}(1), \quad \alpha^{(i+1)\prime}(0) = -\alpha^{(i)\prime}(1),$$
$$u^{(i+1)\prime\prime\prime}(0) = u^{(i)\prime\prime\prime}(1) - 2\bar{L}\left(1 + b_2/b_1 - M/b_1\right)\alpha^{(i)\prime}(1) - 2(b_2 - M)\bar{L}^2/b_1 u^{(i)\prime}(1).$$
(2.9)

Define a 6-component vector for each segment by

$$y_j^{(i)}(x) = \left(u^{(i)}(x), u^{(i)\prime}(x), \alpha^{(i)}(x), u^{(i)\prime\prime}(x), u^{(i)\prime\prime\prime}(x), \alpha^{(i)\prime}(x)\right).$$
(2.10)

With $c^{(i)} = (c_1^{(i)}, c_2^{(i)}, c_3^{(i)}, c_4^{(i)}, c_5^{(i)}, c_6^{(i)})$, the general solution provides

$$y_j^{(i)}(0) = C_{jk}^{(0)} c_k^{(i)}, \quad y_j^{(i)}(1) = C_{jk}^{(1)} c_k^{(i)},$$
(2.11)

where the two $6\times 6$ matrices, $C^{(0)}$ and $C^{(1)}$, which depend on $\lambda$, $\bar{L}$, $q$, and are the same for all segments. They are given in the Supplementary Materials. It follows that $y_k^{(i)}(1) = T_{kp} y_p^{(i)}(0)$, where $T_{kp} = C_{kj}^{(1)} C_{jp}^{(0)-1}$ and, again, the same for each segment. The six conditions in (2.5) and (2.9) connecting the solutions in the two segments coming together at each joint can be expressed as $y_j^{(i+1)}(0) = D_{jk} y_k^{(i)}(1)$, where $D$ is a $6\times 6$ matrix which is the same for all joints and given in the Supplementary Materials. The central relation in forming the eigenvalue equation governing stability used in the following sections is

$$y_j^{(i+1)}(0) = S_{jk} y_k^{(i)}(0), \quad \text{with } S_{jk} = D_{jp} T_{pk},$$
(2.12)

where $S$ is the same for all segments. It depends on the eigenvalue $R\kappa_n$, and the dimensionless parameters $b_1$, $b_2$, and $\bar{L} = L/R$.

## 3. A model problem: The stability of two equilibrium states of a bi-rod

The stability of the straight state of a bi-rod shown on the left in Fig. 3 was analyzed by Leanza et al. (2023). Each of the two rod segments has length $L$ and natural curvature $\kappa_n$. The dimensionless natural curvature specifying stability or lack thereof is $L\kappa_n/2\pi$, and this parameter will be used for the same purpose throughout the present paper because $L$, but not $R$, is the same for all states of the ring. For the straight configuration of the bi-rod the range of stability is (Leanza et al., 2023)

$$-\frac{\sqrt{b_1 b_2}}{2} \leq \frac{L\kappa_n}{2\pi} \leq \frac{\sqrt{b_1 b_2}}{2}, \quad \sqrt{b_1 b_2} = \sqrt{B_1 B_2}/B_3. \tag{3.1}$$

If the length is prescribed, then (3.1) provides the range of the natural curvature for stability, while if the natural curvature is regarded as given, it provides the maximum length for a stable bi-rod. The end points of this range correspond to states at which bifurcation can occur. The state at bifurcation may or may not be stable depending on the initial post-bifurcation behavior. A detailed analysis of the post-bifurcation behavior has not been performed, while the transition behavior of the two equilibrium states of the bi-rod, namely the 1-loop line and the 1-loop 8, under external stimuli, is presented in Figs. S1–S4 in the Supplementary Materials based on the multi-segment Kirchhoff rod model and finite element analysis in Part II.

Throughout this paper, we will present the stability range in the manner of (3.1) with the understanding that nature of the stability at the end values of the range has not been assessed. The bifurcation, or buckling, mode associated with the end values of the range (3.1) for the straight configuration is

$$u_3^{(1)} = u_3^{(2)} = L\left(x - \sin(\pi x)/\pi\right), \quad \alpha^{(1)} = \alpha^{(2)} = \sqrt{B_1/B_2}\sin(\pi x), \tag{3.2}$$

where the upper rod is labeled by (1) and the lower by (2) consistent with the convention in Fig. 2. Here, and throughout the paper, the mode has been normalized such that the maximum value of $u_3/L$ is unity, and rigid body motion is suppressed by requiring $u_3^{(1)} = u_3^{(1)\prime} = \alpha^{(1)} = 0$ at $s = 0$. In this paper, $\alpha$ (in radians) is defined using the right-hand rule for an observer traversing around the loop in sense of increasing $s$ indicated in Fig. 3a. In the earlier paper (Leanza et al., 2023) analyzing straight configurations, $s$ was defined as increasing from the left end of both rods. This accounts for the sign difference between $\alpha^{(2)}$ in (3.2) and that in the previous paper.

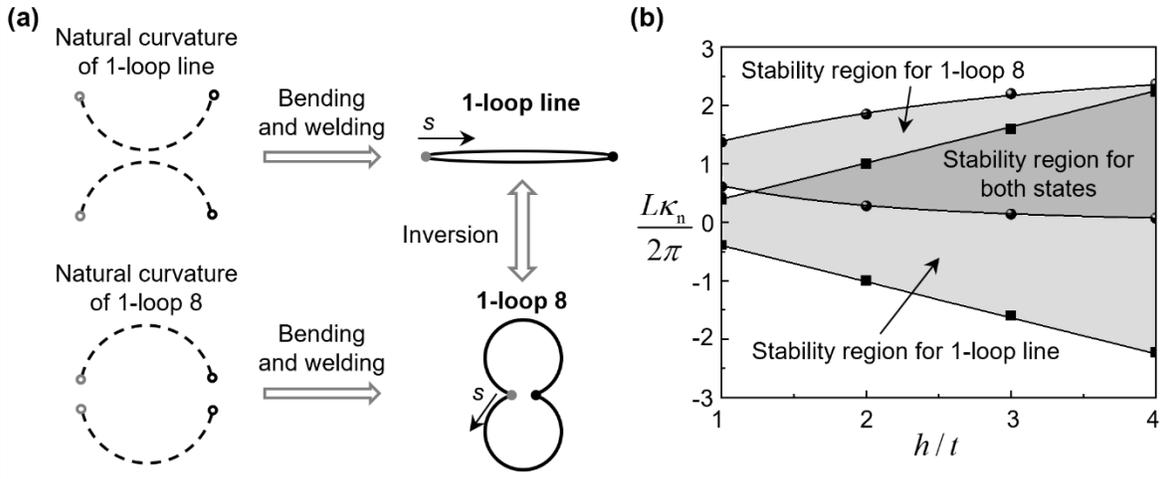

Fig. 3. a) A bi-rod: two rod segments with natural curvature $\kappa_n$ that are straightened and welded at their ends. The dashed curves indicate a positive natural curvature of the detached arcs. The length of each rod is $L$ and the magnitude of the bending moment in each segment is $B_3\kappa_n$. The shape below is an inverted equilibrium state of the bi-rod where the two half-loops segments have radius $R = L/2\pi$ and carry a uniform bending moment of magnitude $B_3(R^{-1} - \kappa_n)$. The flip of the dashed arcs depicting a positive natural curvature from one state to the other occurs due to inversion. The convention for measuring the distance starting at $s = 0$ and the direction for traversing the segments is shown. b) Range of the dimensionless natural curvature, $L\kappa_n / 2\pi$, for stability of the two equilibrium states, 1-loop line and 1-loop 8, of bi-rods having rectangular cross-sections with aspect ratio $h/t$ and Poisson's ratio $\nu = 1/3$, including the range for stability of both states. Independent simulations from Part II are shown as solid dots, and details on how to verify these stability limits using the methods in Part II are provided in Figs. S5 and S6 in the Supplementary Materials.

The quadratic functional governing stability of the 1-loop 8 state of the bi-rod is (2.4) with the summation limited to 2 segments. Periodicity of the solution around the ring requires that $y_j^{(3)}(0)$ is equal to $y_j^{(1)}(0)$. By (2.12), this requires that

$$\left(S_{ik}S_{kj} - I_{ij}\right) y_j^{(1)}(0) = 0, \tag{3.3}$$

where $I$ is the $6 \times 6$ identity matrix. This is the eigenvalue equation for the upper and lower values of $R\kappa_n$ specifying the range of stability of this configuration. Details of the determination of these two eigenvalues are similar to those described in detail for 6-segment hexagrams in the next section and need not be repeated here.

The upper and lower limits of $L\kappa_n/2\pi$ for the straight state, (3.1), and for the 1-loop 8 state of a bi-rod with a rectangular cross-section characterized by an aspect ratio $h/t$ (with Poisson's ratio $\nu = 1/3$) are shown in Fig. 3b. For the rectangular cross-section, $b_1 = B_1/B_3 = (h/t)^2$ and

$$b_2 = \frac{B_2}{B_3} = \frac{2}{1+\nu}\left(1 - \frac{192}{\pi^5}\frac{t}{h}\tanh\left(\frac{\pi}{2}\frac{h}{t}\right)\right). \tag{3.4}$$

The expression for $b_2$ is an approximation given by Sokolnikoff (1956). A bi-rod with a square cross-section cannot be stable in both states. (Analogous results for a bi-rod with circular cross-sections will be included in the next section.) Included in Fig. 3b as solid circular or square dots are numerical results for the stability limits computed using a multi-segment Kirchhoff rod model and finite element analysis given in Part II, which are in excellent agreement with the present approach. For aspect ratios greater than about 1.2, stability in both states is possible, and the range of the natural curvature for mutual stability is indicated in Fig. 3b. Experimental demonstrations of the two states were presented in Leanza et al. (2013), and Video 1 presenting realizations of the two states is included in the Supplementary Materials accompanying Part I. Related investigations of the stability of the multiple sates of a more complex type of bi-rod, referred to as a bigon, have been published by Yu et al. (2021) who carry out a numerical analysis of the stability of the states and transitions between them together with experimental demonstrations.

The bifurcation mode associated with the upper and lower eigenvalue for the 1-loop line state is given by (3.2). The modes for the upper and lower eigenvalues for the 1-loop 8 state are computed numerically, as described in the next section, and plotted in the Fig. 4. The modes for the two states are quite different. The 1-loop line mode produces a displacement of the cusp at the right end relative to the left end. By contrast, the 1-loop 8 mode produces no relative displacement of the cusps at the ends of the two rods. In these modes, the $u_3$ displacement at corresponding points along each of the two rods is the same, while the rotation at those points has the opposite sense (while $\alpha$ is the same, the direction of the observer has reversed and by the right hand rule so does the rotation relative to any fixed direction).

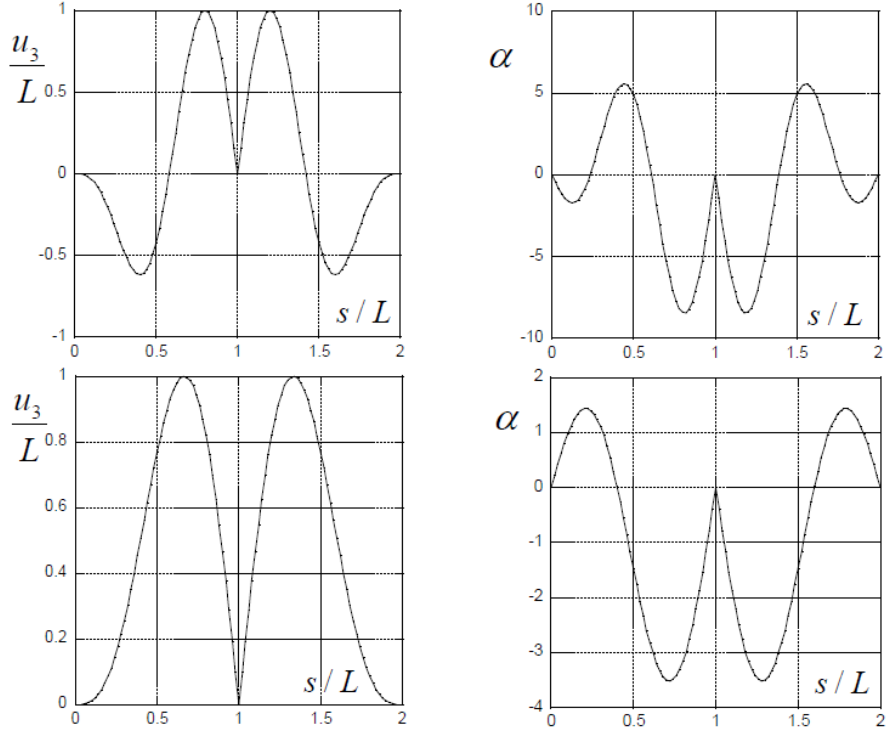

Fig. 4. Bifurcation modes for the limits of the stability range of the 1-loop 8 state of the bi-rod with a rectangular cross-section having $h/t = 2$ and $\nu = 1/3$. The mode associated with the upper eigenvalue is in the upper half of the figure and with the lower eigenvalue in the lower half. The distance $s$ increases around the ring in the manner indicated in Fig. 3 such that $0 < s/L < 1$ is segment (1) and $1 < s/L < 2$ is segment (2). The sense of the rotation $\alpha$ (in radians) is as seen by an observer progressing about the ring using the right-hand rule. The modes are normalized such that the maximum of $u_3/L$ is unity, and rigid body motion is suppressed by requiring $u_3^{(1)} = u_3^{(1)\prime} = \alpha^{(1)} = 0$ at $s = 0$.

## 4. The stability of equilibrium states of the curved hexagram

We begin by considering the four equilibrium states of the curved-sided hexagram in Fig.1: the star, daisy, 3-loop line, and 3-loop 8. Each of these states consists of a uniform bending moment, $B_3(R^{-1} - \kappa_n)$, in each of the six rods, with the total strain energy in the ring being $3B_3 L(R^{-1} - \kappa_n)^2$. The transformation of the star to the daisy, or the 3-loop line to the 3-loop 8, is an inversion in which each cross-section undergoes a $180°$ rotation about its centerline such that the outer surface becomes the inner surface and vice versa. The inversion causes the sense of the natural curvature (the curvature of the detached rod) to flip as well, as illustrated in Fig. 1. To obtain the eigenvalue equation governing bifurcation from any of four these states, denote the 6×6 matrix in (2.12) by $S$ and denote the 6 component vector, $y_k^{(i)}(0)$, by $y^{(i)}(0)$, such that (2.12) becomes $y^{(i+1)}(0) = S\, y^{(i)}(0)$. Continuity around the six-sided ring requires $y^{(7)}(0) = y^{(1)}(0)$, which can be stated as

$$\left(S^6 - I\right) y^{(0)}(0) = 0, \qquad (4.1)$$

where the components of $S$ depend on the eigenvalue, $R\kappa_n = (L\kappa_n/2\pi)/(\bar{L}/2\pi)$, $b_1$, $b_2$, and $\bar{L} = L/R$. Because the rod segments are unstressed when $R\kappa_n = 1$, this value always lies within the stable range. Thus, the upper limit to the stable range is the smallest value of $R\kappa_n$ greater than 1 such that a non-trivial solution to (4.1) exists that is not a rigid body motion. Similarly, the lower limit of the stable range is the largest value of $R\kappa_n$ less than 1 for which a non-trivial solution exists.

The matrix in (4.1) generating the eigenvalues has the structure

$$S^6 - I = \begin{bmatrix} 0 & A^{(1)} \\ 0 & A^{(2)} \end{bmatrix}, \qquad (4.2)$$

where $A^{(1)}$ and $A^{(2)}$ are 3×3 matrices (a similar structure holds for $S^2 - I$ in (3.3)). By (2.10) and (4.2), the first three equations in (4.1) require

$$A^{(1)} \left[ u^{(1)\prime\prime}(0), u^{(1)\prime\prime\prime}(0), \alpha^{(1)\prime}(0) \right] = 0, \qquad (4.3)$$

which, in turn, requires the determinant of $A^{(1)}$ to vanish for any non-trivial solution. This is a necessary, but not sufficient, condition for a non-trivial solution to (4.1) to exist. Let $R\kappa_n$ be any value such that $|A^{(1)}| = 0$, and let $\left[ u^{(1)\prime\prime}(0), u^{(1)\prime\prime\prime}(0), \alpha^{(1)\prime}(0) \right]$ be any associated non-trivial solution to (4.3). To generate a solution to (4.1), the last three equations in (4.1) require

$$A^{(2)} \left[ u^{(1)\prime\prime}(0), u^{(1)\prime\prime\prime}(0), \alpha^{(1)\prime}(0) \right] = 0. \qquad (4.4)$$

In summary, the numerical procedure for determining the upper and lower limits of the range of stability of $R\kappa_n$ is a systematic search for the smallest value of $R\kappa_n$ greater that 1 (and, the largest value of $R\kappa_n$ less than 1) such that $\left|A^{(1)}\right|=0$ and $A^{(2)}\left[u^{(1)''}(0),u^{(1)'''}(0),\alpha^{(1)'}(0)\right]=0$, where $\left[u^{(1)''}(0),u^{(1)'''}(0),\alpha^{(1)'}(0)\right]$ is a non-trivial solution to (4.3). Modal rigid-body displacement and rotation is eliminated by requiring $\left[u^{(1)}(0),u^{(1)'}(0),\alpha^{(1)}(0)\right]=0$. Thus, the associated mode has

$$y^{(1)}(0) = \left[0,0,0,u^{(1)''}(0),u^{(1)'''}(0),\alpha^{(1)'}(0)\right], \tag{4.5}$$

and the set of initial values $y^{(k)}(0)$ in each of the rod segments is obtained sequentially using $y_j^{(i+1)}(0) = S_{jk} y_k^{(i)}(0)$. Within each segment, the solution is given by (2.7) or (2.8) with

$$c_j^{(i)} = C_{jk}^{(0)-1} y_k^{(i)}(0).$$

In Leanza et al. (2023) it was *incorrectly* determined that the upper and lower limits of the stable range of $L\kappa_n / 2\pi$ for the 3-loop line state of the hexagram are the same as those for the 1-loop line state of the bi-rod given by (3.1). The correct stability range for the 3-loop line state of the hexagram is

$$-\frac{\sqrt{b_1 b_2}}{6} \leq \frac{L\kappa_n}{2\pi} \leq \frac{\sqrt{b_1 b_2}}{6}. \tag{4.6}$$

The incorrect range published earlier was a consequence of overlooking eigenvalue solutions closer to $R\kappa_n = 1$ than the limits given by (3.1). It should be noted that results for both the line state of the bi-rod and for the 3-loop line state of the hexagram were first obtained numerically. The analytical formulas given in (3.1) and (4.6) were 'guessed' with help of insight from analytical formulas for a single rod given in Leanza et al. (2023). Subsequent numerical calculation established these formulas as being accurate to better than $10^{-7}$ for all values of the parameters checked. Thus, we feel it is reasonable to anticipate that (3.1) and (4.6) are, in fact, exact results.

The evaluation of the stability range for the other states of the curved hexagram is carried out numerically as outlined above, as is the computation of the associated bifurcation modes. Noting again that the dimensionless parameters that must be assigned are $b_1$, $b_2$ and $\bar{L} = L/R$, it can be seen that the only parameter change from one state to the other of a curved hexagram is $\bar{L} = L/R$, and this also holds for the two additional states revealed in Section 4.3. For the star state, $L/R = 2\pi/3$, for the daisy, $L/R = 4\pi/3$, and for the 3-loop 8, $L/R = 2\pi$. The formulation also applies to the 3-loop line with $L/R = 0$, noting that $M\bar{L} = L\kappa_n$, but the general solution in (2.7) and (2.8) has a different form, which is given in the earlier paper (Leanza et al., 2023). It should also be noted that in the earlier paper, $s$ was taken to increase starting at the left end of each segment rather than having $s$ increase as the ring is traversed from segment to segment. This accounts for the sign difference in the modal $\alpha$ in the two papers for the even numbered rod segments.

*4.1. Stability range of the natural curvature*

We begin by presenting in Fig. 5 the stability range of $L\kappa_n/2\pi$ for two states of the bi-rod and four states of the curved hexagram for the case in which their rod segments have circular cross-sections with $\nu = 1/3$. These results, for which $b_1 = 1$ and $b_2 = 1/\sqrt{1+\nu}$, apply to solid circular cross-sections and to annular cross-sections bounded by concentric circular rings. The stability range of the two states of the bi-rod does not overlap, and thus for rods with circular cross-sections it is not possible for a bi-rod to be stable in both states. The limited set of experiments carried out in Leanza et al. (2023) bore this out. For the curved hexagon, the range of the dimensionless natural curvature, $L\kappa_n/2\pi$, is given by Eq. (4.6) for the 3-loop line state, by (0.192, 0.477) for the star state, by (0.524, 0.816) for the daisy state, and by (0.857, 1.286) for the 3-loop 8 state. As seen in Fig. 5, there is no overlap of the stability ranges of any of the four states of the curved hexagram. Thus, for bi-rods or the curved-sided hexagram constructed from rods with circular cross-sections, it is only possible to fabricate an entity that has one of the states considered here to be stable. Of course, that entity could be deformed and constrained so that it remained in one of the unstable states, but as soon as the constraint was removed, it would snap to the stable state, or possibly another state that has not been considered.

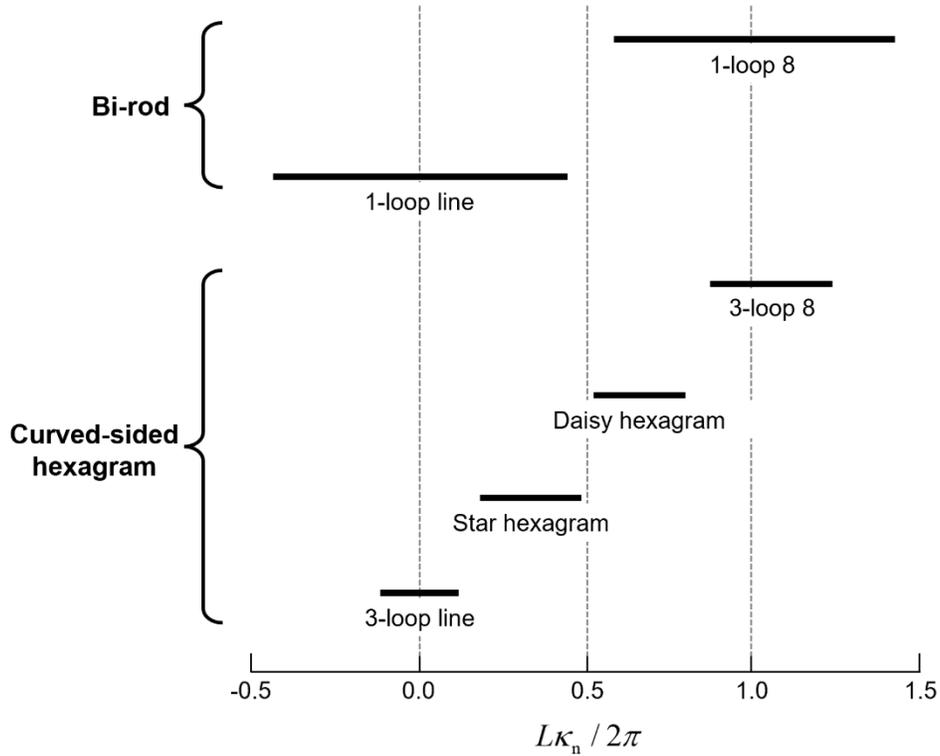

Fig. 5. Stability ranges of $L\kappa_n/2\pi$ for two states of the bi-rod and four states of the curved-sided hexagram for rod segments with circular cross-sections ($\nu=1/3$) revealing that there is no overlap of the ranges of stability.

Fig. 6 presents the upper and lower limits of the stability range of $L\kappa_n/2\pi$ for each of the four states of the curved-sided hexagram with rod segments having a rectangular cross-section with $\nu=1/3$. The horizontal dashed line indicates the value of $L\kappa_n/2\pi$ for which the bending moment in rod segments in the basic state is zero, i.e., for which $R\kappa_n=1$, or, for the line states, $\kappa_n=0$. As noted earlier, when the bending moment in the basic state is zero, that state is stable. Also included in Fig. 6 are results computed by a multi-segment Kirchhoff rod model and finite element analysis presented in Part II and denoted as heavy dots. The results are in excellent agreement. The stability limits for the 3-loop line and 3-loop 8 states of the hexagram lie within those for the corresponding 1-loop states for the bi-rod even though, apart from the number of loops, the parameters governing them are the same. Some perspective on this can be obtained by noting that the eigenvalues of the bi-rod are given by $(S^2-I)y=0$ while those of the hexagram are given by $(S^6-I)y=0$. Because $(S^6-I)=(S^4+S^2+I)(S^2-I)$, it follows that any eigenvalue of the 1-loop line (or 1-loop 8) of the bi-rod is an eigenvalue of the 3-loop line (or 3-loop 8) of the hexagram, because $S$ is the same for the two line states (and the two loop 8 states). The converse does not necessarily hold, and, indeed, each 3-loop state of the hexagram has an eigenvalue range that is smaller than that of the respective 1-loop state of the bi-rod.

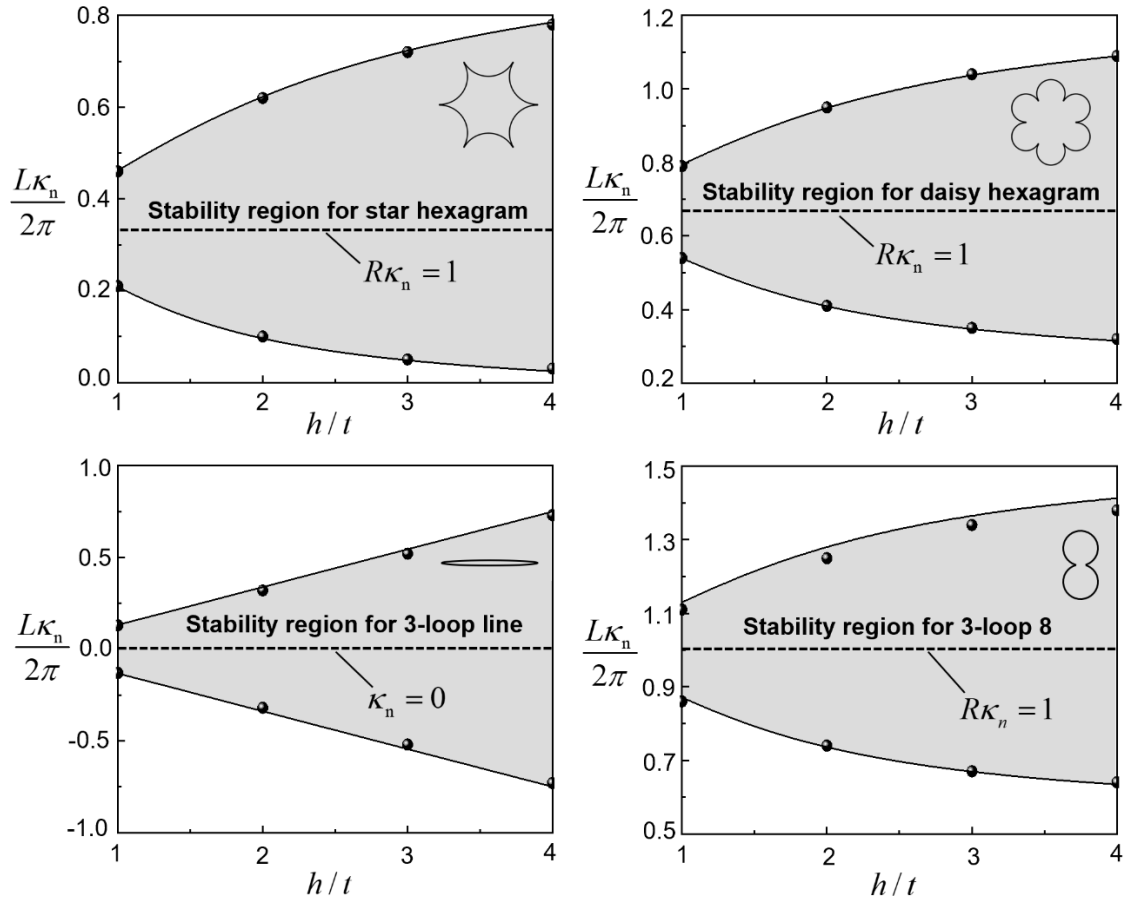

Fig. 6. Upper and lower limits of the stability range of $L\kappa_n/2\pi$ for four states of the curved hexagram for rods with a rectangular cross-section having aspect ratio $h/t$ and $\nu=1/3$. The horizontal dashed line indicates the value of $L\kappa_n/2\pi$ which the bending moment in the basic state vanishes. The solid dots are results for the stability limits computed by the multi-segment Kirchhoff rod model and finite element simulations given in Part II. Details on how to verify the stability limits for the four equilibrium states using the methods in Part II are provided in Figs. S7–S10 in the Supplementary Materials.

If the cross-section is square ($h/t = 1$), there is again no overlap of any of the four basic states. As was the situation for circular cross-sections, the star and daisy states, which are inversions of one another, cannot both be stable. (The higher relative torsional stiffness of the circular cross-section compared to the square cross-section (i.e., $b_2 = 0.866$ compared to $b_2 = 0.637$ for $\nu = 1/3$) expands the stability range for circular cross-sections but not enough to render the star and daisy states mutually stable). Loss of stability in any of the states considered takes place in a mode having out-of-plane bending coupled with twist, and thus stability is enhanced by relative increases in the out-of-plane bending stiffness and the torsional stiffness, as the formulas for the straight states, (3.1) and (4.6), indicate. For rectangular cross-sections, the major factor in enhancing the stability range is increase in the out-of-plane bending stiffness. Stability of all four states becomes possible when $h/t > 3.4$, with the 3-loop line and 3-loop 8 states setting the limits of the mutual stability as plotted in Fig. 7. A more detailed version of this figure is presented in Fig. S11 in the Supplementary Materials showing the ranges of mutual stability of sub-sets of the four states.

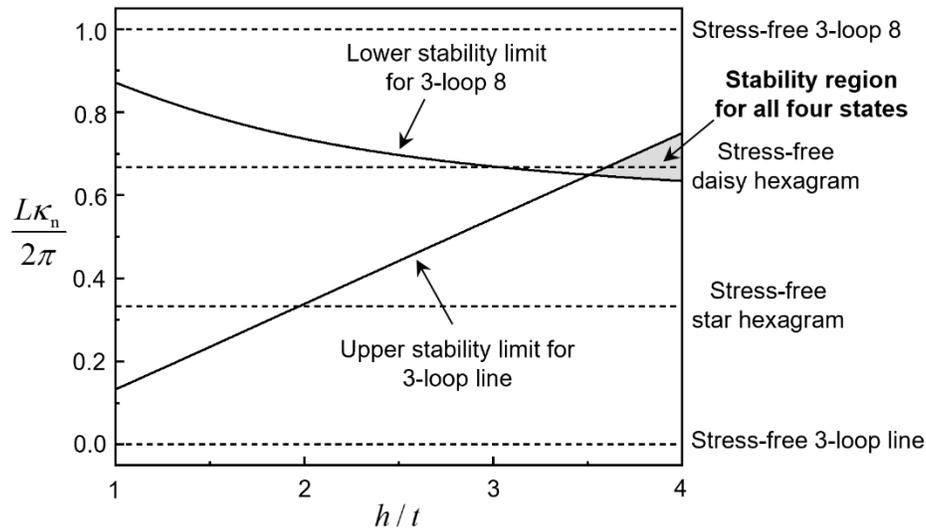

Fig. 7. For the curved-sided hexagram, all four equilibrium states (star, daisy, 3-loop line and 3-loop 8) are stable in the shaded area indicated for rod segments of rectangular cross-section with $h/t > 3.4$ ($\nu = 1/3$). In this range, the upper stability limit is set by the upper limit of the 3-loop line state and the lower limit by the lower limit of the 3-loop 8 state. The value of the dimensionless natural curvature associated with stress-free conditions of each of the states is shown as a dashed line.

The lesson to be learned from this section is that some control over the stability of the multitude of equilibrium states of the curved hexagram is possible by judicious choices of the natural curvature and the cross-section of the rod segments. This is illustrated experimentally in Video 2 in the Supplementary Materials accompanying Part I for three curved stainless-steel hexagrams having rectangular cross-sections ( $h/t = 4$ ) and different natural curvatures. Experimental details on how to fabricate these rings are provided in Appendix B. For the first hexagram, $L\kappa_n / 2\pi$ is in the range (0.03, 0.32) and is seen to share two stable states, 3-loop line and star. For the second hexagram, $L\kappa_n / 2\pi$ is in the range (0.32, 0.63), and it has three stable states, the 3-loop line, star, and daisy. The third hexagram, with $L\kappa_n / 2\pi = 2/3$, corresponds to a stress-free daisy state and is stable in all four of the states, as can be seen from Fig. 7. These experimental examples are in accord with the stability ranges in Figs. 6 and 7. The first half of Video 3 demonstrates the transitions from one state to another for the curved hexagram having the four mutually stable states. Fig. 8 presents experimental realizations of the four states of the curved hexagram taken from the videos, and it also includes realizations of the two 6-circle states introduced in Section 4.3.

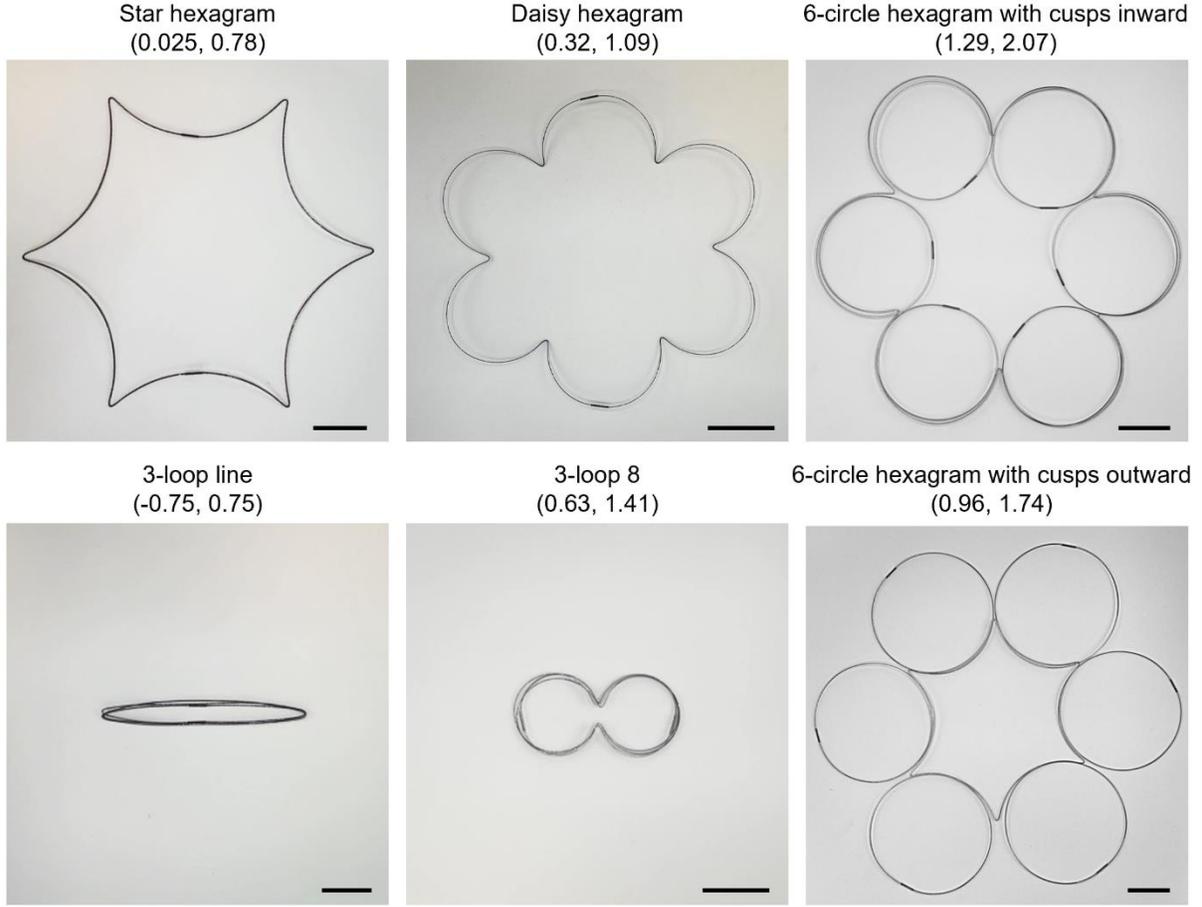

Fig. 8. Physical realizations of the four states in Fig. 1 and the two 6-circle states in Section 4.3 taken from Videos 2 and 3 in the Supplementary Materials. The numbers in brackets denote the dimensionless natural curvature range within which the state is stable. All states have a rectangular cross-section with an aspect ratio $h/t$=4 and Poisson's ratio $v$=1/3. The four states in Fig. 1 have an edge length $L$=200mm, which are fabricated by connecting two 600mm stainless-steel rods. The two 6-circle states in Section 4.3 have an edge length $L$=600mm, which are fabricated by connecting six 600mm stainless-steel rods. Scale bars: 5cm.

*4.2. A small selection of bifurcation modes associated with the eigenvalue limits*

In the two figures of the modes to follow, the modes are normalized such that the maximum of $u_3/L$ is unity, and rigid body motion is suppressed by requiring $u_3^{(1)} = u_3^{(1)\prime} = \alpha^{(1)} = 0$ at $s=0$. In Fig. 9 for the upper limit of the 3-loop 8 for rods having a rectangular cross-section with $h/t=2$, one sees that $u_3$ is symmetric about the central cusp at $s/L=3$ joining rods 3 and 4. The largest modal defections occur in rods 2 and 5, the deflection at the central cusp is zero, and the deflections at the cusps at ends of rods 1, 2, 4 and 5 are equal. Because $s$ is taken to increase traversing the ring and the right-hand rule defines a positive rotation, the fact that $\alpha$ is symmetric with respect to the central cusp means that the rotation relative to a single fixed axis parallel to that cusp is anti-symmetric.

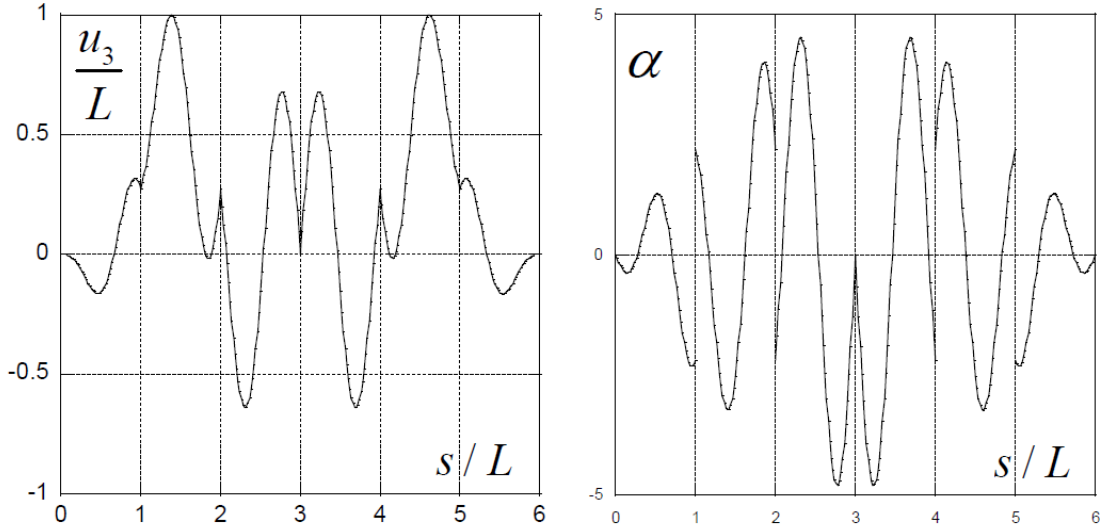

Fig. 9. The bifurcation mode associated with the eigenvalue determining the upper limit of stability for the 3-loop 8 state of the curved hexagram for rods with $h/t = 2$ and $v = 1/3$. The rotation is as seen by an observer traversing the ring using the right-hand rule and thus the rotation undergoes a change in sign across each cusp, as does the slope of the deflection.

An example presenting the bifurcation mode associated with the eigenvalue determining the upper stability limit of the star state of the curved hexagram is given in Fig. 10. There is again symmetry of $u_3$ and $\alpha$ about the central cusp of the ring at $s/L = 3$, and anti-symmetry of rotation about a single fixed axis parallel to the cusp. Now, the largest deflection occurs at the central cusp. The bifurcation motion is an upward motion of the star generally increasing towards the central cusp accompanied by rod rotations (with respect to the fixed $x_1$-axis) that have opposite signs above and below $x_2 = 0$ at corresponding locations.

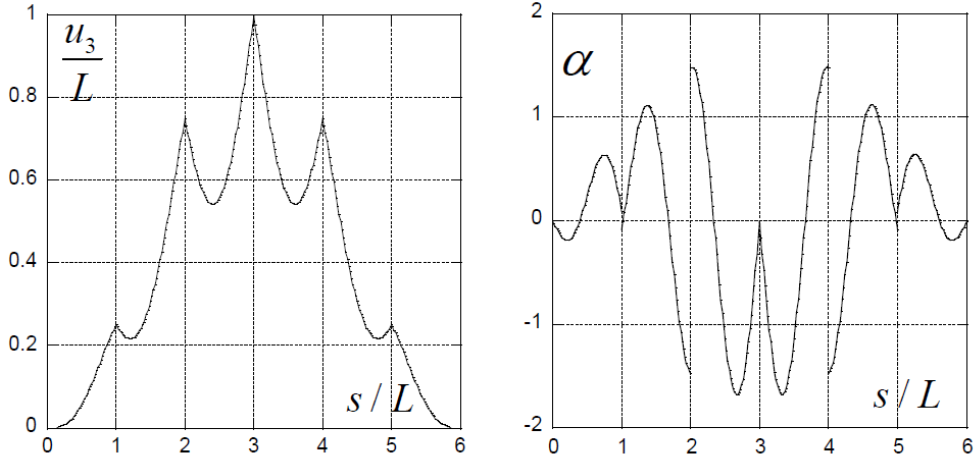

Fig. 10. The bifurcation mode associated with the eigenvalue determining the upper limit of stability for the star state of the curved hexagram for rods with $h/t=2$ and $v=1/3$. The rotation is as seen by an observer traversing the ring using the right-hand rule and thus the rotation undergoes a change in sign across each cusp, as does the slope of the deflection.

*4.3. Two 6-circle hexagrams*

In carrying out a finite element analysis of the transitions between equilibrium states using a method employed in Part II for curved-sided hexagrams with rectangular cross-sections and high natural curvature, we discovered several previously unobserved states, two of which are seen in the inserts of Fig. 11: a 6-circle hexagram with cusps pointing inward and another with cusps pointing outward. Each of the rod segments of the 6-circle hexagram with cusps pointing inward is a circle with radius $R=3L/10\pi$ such that the rod overlaps itself over 2/3 of its circular circumference—the dark arc segments in Fig. 11 indicate overlap. The rod segments of the 6-circle hexagram with cusps pointing outward have $R=3L/8\pi$ and overlap over 1/3 of their circumference. As for the other equilibrium states investigated here, each rod segment of a 6-circle hexagram supports the same uniform bending moment. Transition between the previously discussed states of the curved hexagram and either of the two configurations is complicated, as is the transformation between the two states in Fig. 11 resulting in an inversion. Curiously, the overlap regions of the segments have a daisy shape when cusps point inward and a star shape when cusps point outward. The stability analysis described earlier applies to these new configurations with $L/R=10\pi/3$ for cusps in and $L/R=8\pi/3$ for cusps out. The stability limits for rod segments with rectangular cross-sections for the two cases are plotted in Fig. 11.

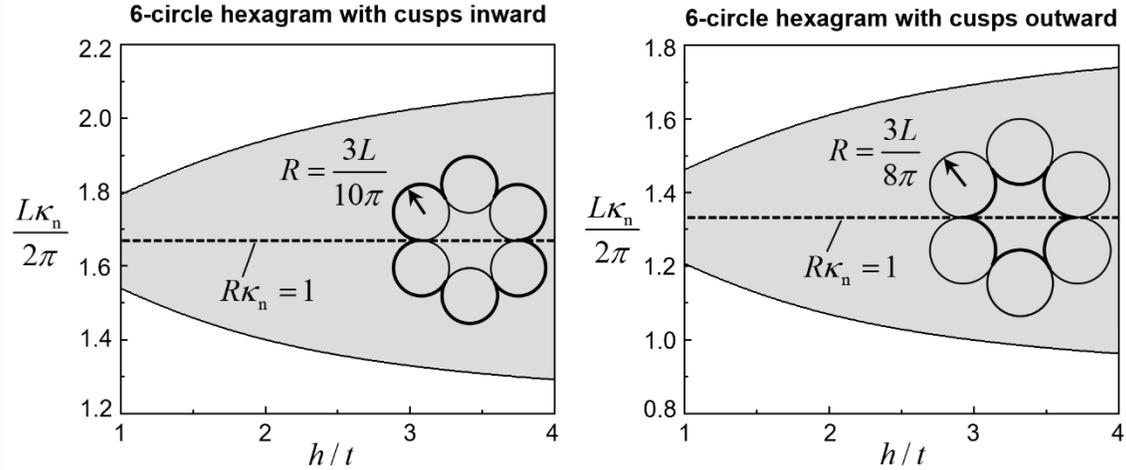

Fig. 11. Stability limits for 6-circle hexagrams, with cusps pointing inward on the left and cusps pointing outward on the right, for rod segments with rectangular cross-sections ($v = 1/3$). In the inserts, the portion of the circumference for each rod overlapping itself is indicated by the heavy arc, covering 2/3 of each circle for cusps inward and 1/3 of each circle for cusps outward.

When the cusps point inward, there is no overlap with the stability range of any of the four states plotted in Fig.6 in the range $h/t < 4$. However, for $h/t \geq 4$ and $L\kappa_n/2\pi \cong 1.3$, the 6-circle state with cusps inward is mutually stable with the 3-loop 8 state. When the cusps point outward, there is a range of the aspect ratio plotted such that the 6-circle state is mutually stable with the daisy state and the 3-loop 8 state, i.e., approximately, for $h/t > 3$ and $L\kappa_n/2\pi \cong 1$. When the cross-section is square, there is no natural curvature for which the two 6-circle states are mutually stable, but for rectangular cross-sections with $h/t$ greater than about 1.5, a mutual range exists which is roughly centered about $L\kappa_n/2\pi = 1.5$. The second half of Video 3 in the Supplementary Materials includes an experimental demonstration of a curved-sided hexagram with properties chosen to fall within the range of mutual stability ($h/t = 4$ and $L\kappa_n/2\pi = 1.5$) which is seen to invert back and forth between the two stable 6-circle states.

There are infinitely many equilibrium states in the form of the 6-circle hexagrams (with an increasing number of loops in each circle) that are all planar and all have uniform bending in the equilibrium state. In this paper, our emphasis has been on the stability of the planar, uniform bending equilibrium states, and we have not conducted a systematic search for stable or unstable nonplanar states. However, we expect stable nonplanar equilibrium states must exist. For example, consider the curved-sided hexagram with circular or square cross-section (c.f., Figs. 5 and 6). There are no mutually stable states among the planar states considered. Suppose the natural curvature of a hexagram fell within one of the instability gaps between the stable states. It seems obvious that there must be at least one stable state for that hexagram. That state is likely to be nonplanar with nonuniform bending and twist. In Fig. S12 in the Supplementary Materials, we have verified the existence of one nonuniform nonplanar stable state using the rod model introduced in Part II for a hexagram with square cross-section ($h/t = 1$) having a dimensionless natural curvature $L\kappa_n / 2\pi = 0.50$. It can be seen from Fig. 6 that this natural curvature lies in the instability gap between stable star and daisy states. The shape of the hexagram in the stable nonplanar state has the appearance of a star that is curling out of its plane towards the inverted daisy shape, as shown in Fig. S12 in the Supplementary Materials.

## 5. Concluding remarks

Parts I and II of this paper investigate the multiple equilibrium states of a special elastic curved-sided hexagram with multiple equilibrium states including a collapsed 3-loop line state. Part I employs the classical criterion based on the second variation of the system energy to establish the parameter range for stability of the curved hexagram for each of six of the basic states considered. The method employed is a specialized version of Kirchhoff rod theory, derived by Leanza et al. (2023), which is shown to be relatively straightforward to implement and effective for stability investigations. Experimental demonstrations of stable and unstable states are presented that are in good agreement with the theoretical predictions. Guided by the theoretically predicted stability ranges for the basic states, a curved hexagram has been designed and tested which has four mutually stable states, including the 3-loop collapsed line state. Experimental demonstrations of the transitions among the various states are shown in three videos accompanying Part I in the Supplementary Materials. In these videos, and in Part II, it is seen that the transitions between the various states take place without the occurrence of self-contact between the six rod segments comprising the curved hexagram. Part II will focus on details of the transitions between states, including more extensive experimental illustrations of the phenomena, and it will present an alternative method for assessing stability of the basic states analyzed in Part I.

## Acknowledgments

R.R.Z., L.L, J.D., S.L. acknowledge National Science Foundation Award CPS-2201344 and National Science Foundation Career Award CMMI-2145601 for the support of this work.

## Supplementary materials

Supplementary material associated with this article can be found in the online version.

## Appendix A. Side views of the curved-sided hexagrams

Fig. A1 shows experimental side view images of the four basic equilibrium states of the curved-sided hexagram. It is seen that they stay in plane in the equilibrium state without out-of-plane deformation.

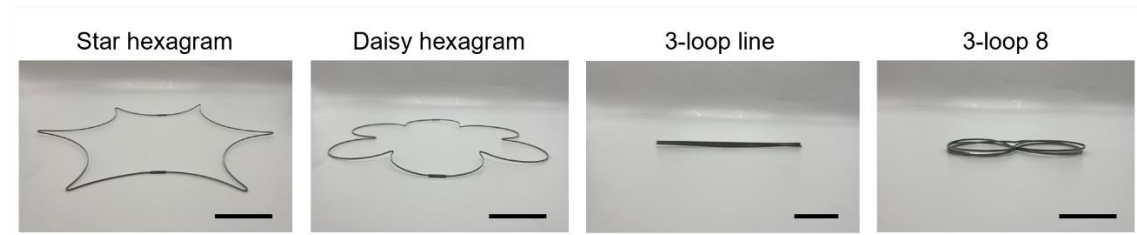

Fig. A1. Side views of the four basic equilibrium states of the curved-sided hexagram. Scale bars: 6cm.

## Appendix B. Fabrication of the curved-sided hexagram rings

The curved-sided hexagram rings presented in this work are fabricated by manually reshaping stainless-steel rods. In the experiments, connections between adjacent curved segments are achieved by plastically deforming a long rod, which in turn introduces small "corners" to the rod. When the corner size is relatively small, such as the 1 mm corner radius used in Part I and Part II, it has negligible effects on the stability of the rings. This negligible effect of the corner radius is demonstrated in Figs. 3 and 6, where the stability ranges obtained using the numerical method in Part II agree with the theoretical predictions very well. The four basic equilibrium states (i.e., the star hexagram, the daisy hexagram, the 3-loop line, and the 3-loop 8) shown in Fig. 8 all have the same edge length, $L = 200$ mm, and are fabricated using two 600 mm long stainless-steel rods with rectangular cross-section of height $h = 2$ mm and thickness $t = 0.5$ mm. Fig. A1 shows the fabrication process of the star hexagram as an example. First, we reshape the two long rods into multiple curved-sided segments by applying plastic deformations. Each long rod has three corners. Then, we connect the two rods using two joints to obtain the star hexagram. Note that the natural curvature of each segment can also be manually controlled by applying plastic deformation. The two 6-circle hexagram rings are fabricated using the same method, except that six 600 mm stainless-steel rods were used (each with only one corner) in order to achieve the desired natural curvature. Note that both of the 6-circle hexagram rings can be deployed to a star hexagram state of edge length $L = 600$ mm, however, the obtained star hexagram state is unstable and will promptly snap to a 6-circle hexagram state once external constraints are removed.

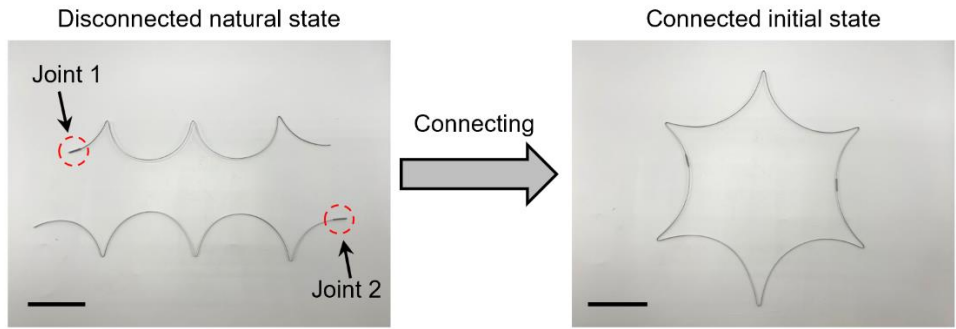

Fig. A2. Fabrication process of the star hexagram by connecting two long stainless-steel rods. Scale bars: 8cm